\documentclass[prb,twocolumn,amsmath,amssymb,floatfix]{revtex4}

\usepackage{graphicx}
\usepackage{epstopdf}
\usepackage{dcolumn}
\usepackage{bm}
\usepackage{color}
\usepackage{natbib}
\usepackage{amsmath}
\usepackage{amssymb}
\usepackage[caption=false]{subfig}

\begin{document}

\title{Why Are There So Few Perovskite Ferroelectrics?}
\author{Nicole A.\ Benedek}
\email{nicole.benedek@austin.utexas.edu}
\affiliation{Materials Science and Engineering Program, The University of Texas at Austin, 1 University Station, Austin, Texas 78712 USA}
\author{Craig J.\ Fennie}
\affiliation{School of Applied and Engineering Physics, Cornell University, Ithaca, New York 14853 USA}
\email{fennie@cornell.edu}

\begin{abstract}
We use a combination of symmetry arguments and first-principles calculations to explore the connection between structural distortions and ferroelectricity in the perovskite family of materials. We explain the role of octahedral rotations in suppressing ferroelectricity in these materials and show that, as the tolerance factor decreases, rotations alone cannot fully suppress ferroelectricity. Our results show that it is cation displacements (`hidden' in Glazer notation) that accompany the rotations, rather than the rotations themselves, that play the decisive role in suppressing ferroelectricity in these cases. We use the knowledge gained in our analysis of this problem to explain the origin of ferroelectricity in $R3c$ materials such as FeTiO$_3$ and ZnSnO$_3$ and to suggest strategies for the design and synthesis of new perovskite ferroelectrics. Our results have implications not only for the fundamental crystal chemistry of the perovskites but also for the discovery of new functional materials.   
\end{abstract}

\maketitle

\begin{figure*}
\centering
\includegraphics[width=9.5cm]{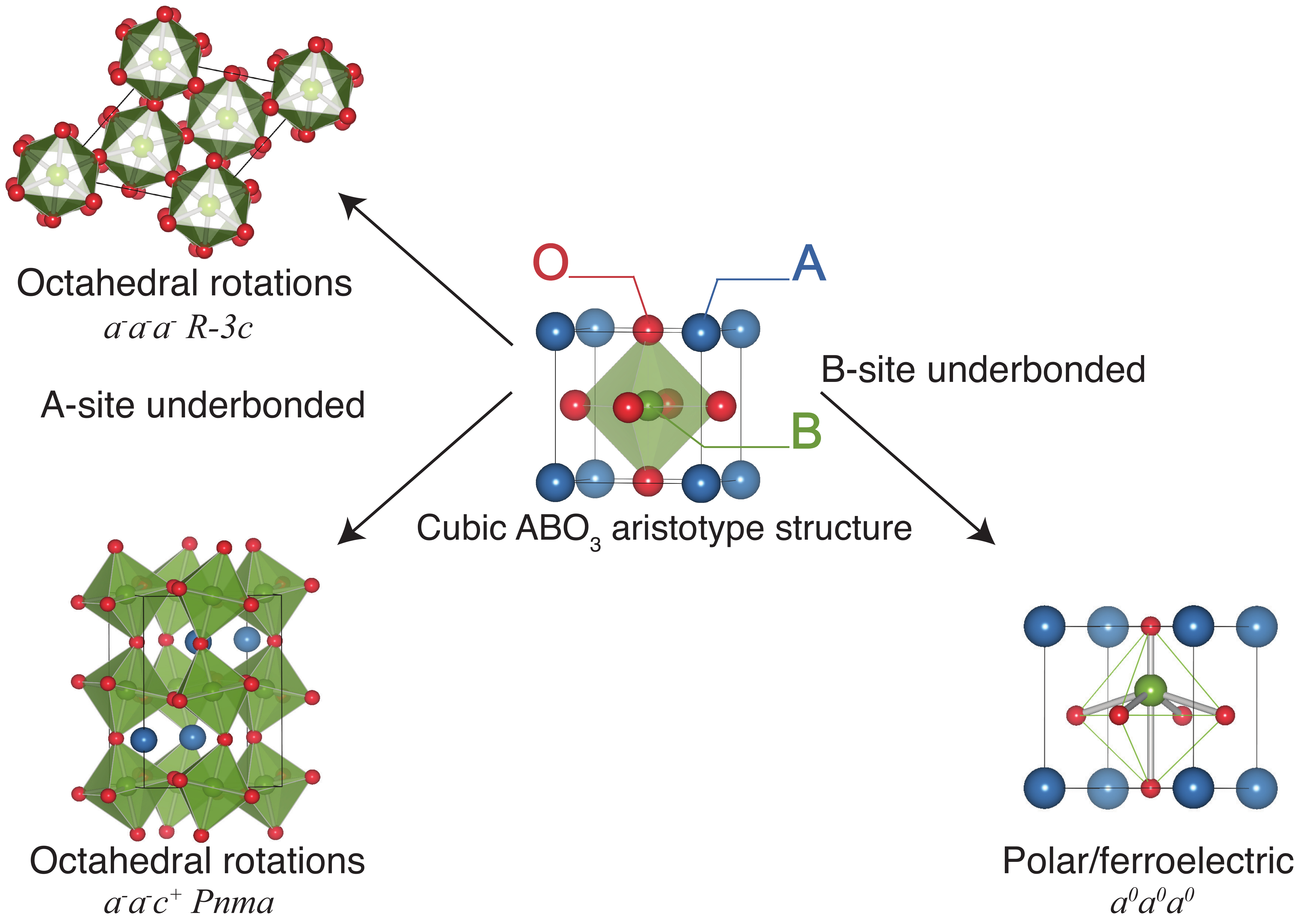}
\caption{\label{cubic}Cubic $Pm\bar{3}m$ perovskite structure and common types of structural distortions. For the rotation distortions, we have shown the two most commonly occurring space groups for perovskites, together with the corresponding Glazer tilt pattern (see text). The A-site cations have been omitted from the $R\bar{3}c$ structure for clarity.}
\end{figure*}

Perovskite oxides are perhaps the most widely studied and technologically important of all the ABO$_3$ phases.\cite{muller74} The remarkable versatility of the perovskite structure (the A and B site can accommodate nearly every element of the periodic table\cite{schlom08}) leads to a huge range of properties, including (but not limited to) ferroelectricity, ferromagnetism and colossal magnetoresistance, piezoelectricity, multiferroicity and metal-insulator transitions. One reason for this is that nearly all cubic perovskites are unstable to energy-lowering structural distortions\cite{rabe07} and hence typically have rich structural phase diagrams. The most common distortions are those that give rise to ferroelectricity (usually an off-centering of the B-site cation) and tilts or rotations of the BO$_6$ octahedra, as shown in Figure \ref{cubic}. 

One of the longest-standing puzzles in the solid-state chemistry of perovskites has been the seeming incompatibility between ferroelectricity and octahedral rotations: prototypical ferroelectrics, such as BaTiO$_3$ and PbTiO$_3$, do not have structures with octahedral rotations, whereas perovskites with octahedral rotations tend not to be polar or ferroelectric. Indeed, most perovskites crystallize in non-polar space groups with octahedral rotations, which serve to optimize the A-site cation coordination environment\cite{woodward97,woodward01} (in contrast, off-centering ferroelectric distortions are driven by B-site bonding preferences). The most commonly occurring space group among perovskites is (non-polar) $Pnma$,\cite{woodward01} shown in Figure \ref{cubic}. However, as we will discuss later, many perovskites with the $Pnma$ structure are unstable to \emph{both} ferroelectric and octahedral rotation distortions in their cubic phases at $T=0$. This unique insight, facilitated by theory, raises the question: why aren't these materials ferroelectric?

In this Feature Article, we reveal why ferroelectricity is suppressed in the $Pnma$ structure and explore the general question of whether and how rotations and ferroelectricity compete in ABO$_3$ materials in general. This question has been the subject of considerable research for some time, in both the chemistry\cite{thomas94,thomas96} and physics\cite{vanderbilt95} communities. Although detailed studies have been performed on individual materials\cite{eklund09,amisi12}, so far there exists no systematic and unifying first-principles theory that is generally applicable to all ABO$_3$ materials. Using density functional theory, in combination with crystal chemistry and symmetry arguments, we present a surprising and counter-intuitive finding: as the tendency of a given material to undergo octahedral rotation distortions increases, rotations by themselves do not generally suppress ferroelectricity, in $Pnma$ or in other space groups that allow rotation distortions, as is commonly thought. It is the A-site anti-polar displacements, induced by the rotations in the $Pnma$ structure, that play a crucial role in suppressing ferroelectricity and in stabilizing the $Pnma$ structure. We show that in the absence of these A-site distortions, most perovskites would adopt the $R3c$ structure and would be ferroelectric (or polar).%

We use the knowledge gained in our analysis of the $Pnma$ structure to suggest design rules for the creation of new A-site driven geometric ferroelectrics,\cite{fennie05,spaldin04,ederer06,singh04,singh06} for which no approaches currently exist. In fact, the most fertile place to search for new ferroelectrics may be the place that has thus far been considered the least likely to contain them: materials that are expected to have large rotations. Geometric ferroelectricity does not involve a chemical mechanism, such as a second-order Jahn-Teller distortion or lone pair active cations and therefore provides greater flexibility in the search and design of materials that combine ferroelectricity with other electronic properties, such multiferroicity, for example.\cite{hill00} Using these principles we explain the origin of ferroelectricity in the recently synthesized $R3c$ polymorphs of the Ilmenite family, such as ZnSnO$_3$ and FeTiO$_3$, and suggest several new targets for synthesis. Our work suggests that although BaTiO$_3$ was the first known perovskite ferroelectric and has perhaps been the most intensively studied, it appears to be an exception rather than the rule and is not representative of the majority of perovskite and perovskite-like ferroelectrics. We provide new insights into the crystal chemistry of the perovskites and strategies for the synthesis of new materials.

\section*{Crystallographic Analysis of Structural Distortions}
\begin{figure}
\centering
\includegraphics[width=8cm]{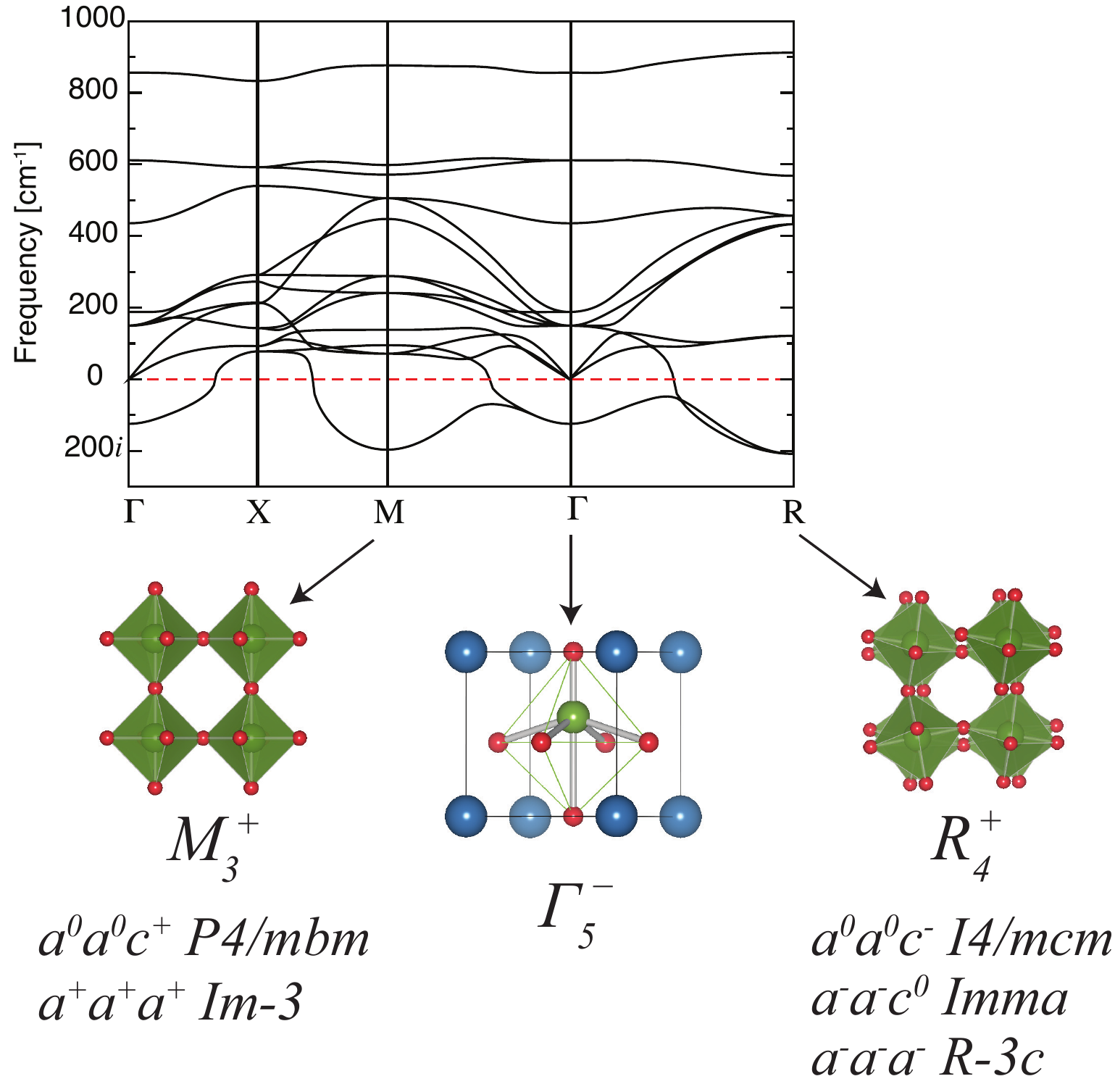}
\caption{\label{catio3_phonons} Phonon dispersion curve of $Pm\bar{3}m$ CaTiO$_3$ from density functional perturbation theory calculations. There are instabilities at three points in the Brillouin zone: $M$, $R$ and $\Gamma$. The $M$ and $R$ instabilities correspond to octahedral rotations whereas the $\Gamma$-point instability corresponds to a polar/ferroelectric distortion. For the $M$ and $R$ distortions, we have listed the most commonly occurring space groups for each, together with their Glazer tilt patterns. Note that the A-site is not fixed by symmetry in $Imma$ and that there are two distinct sites for the A cation in $Im\bar{3}$. The ground state $Pnma$ structure of CaTiO$_3$ is built from a combination of the $a^0a^0c^+$ and $a^-a^-c^0$ rotation patterns.}
\end{figure}

We use the symmetry-mode approach extensively applied by Stokes, Hatch, Perez-Mato and their co-workers\cite{stokes91,hatch01,perez-mato10} to describe structures and structural distortions. We will briefly describe the details relevant to the work presented in this article and refer the reader to the cited references for additional information. Within the symmetry-mode approach, a distorted structure is related to a higher-symmetry `parent' phase by one or more \emph{modes}: atomic displacement patterns that possess specific symmetry properties. For example, if we take $Pm\bar{3}m$ as the parent structure, the distorted structure $R\bar{3}c$ can be related to $Pm\bar{3}m$ by an octahedral rotation mode that transforms like the irreducible representation $R_4^+$ (using the notation of Miller and Love\cite{miller67}). As another example, the polar group $R3c$ can be related to the parent structure $R\bar{3}c$ by a ferroelectric mode that transforms like the irreducible representation $\Gamma_2^-$. Alternatively, we can take $Pm\bar{3}m$ as the parent structure for $R3c$, in which case there are two different modes that connect the space groups: an octahedral rotation mode and a ferroelectric mode.\cite{note_modes} Figure \ref{catio3_phonons} shows the most common distortions encountered in cubic perovskites, their irreducible representation and associated possible space groups. The octahedral rotation modes in ABO$_3$ perovskites are uniquely determined by symmetry and are therefore independent of the composition of a particular material. The same cannot be said of the ferroelectric modes, which do depend on chemistry. We obtained the atomic displacements corresponding to the ferroelectric modes for each material we studied by using density functional perturbation theory to calculate the force-constants matrix, $C$,  
\begin{equation}
C^{\alpha\beta}_{ij}=\frac{\partial^2E}{\partial u_{i\alpha}\partial u_{j\beta}},
\end{equation}
where $E$ is the energy and $u_{i\alpha}$ is the displacement of ion $i$ ($j$) in the Cartesian direction $\alpha$ ($\beta$) from its equilibrium position in the high-symmetry parent structure. The eigenvalues of $C$ are force-constants and the eigenvectors are atomic displacements that correspond to modes of the parent structure. B-site dominated ferroelectrics are those for which the B-site cation makes the largest contribution to the ferroelectric eigenvector, \textit{i.e.}, the B-site cation makes the largest displacement away from the $Pm\bar{3}m$ phase. Vibrational frequencies were obtained by forming and diagonalizing the dynamical matrix, $D$, 
\begin{equation}
D_{ij}^{\alpha\beta} = (M_iM_j)^{-\frac{1}{2}}C^{\alpha\beta}_{ij},
\end{equation}
where $M_i$ and $M_j$ are the masses of ions $i$ and $j$. The eigenvalues of $D$ are the squares of the vibrational frequencies. When we speak of `instabilities' or `unstable' modes, we are referring to modes for which the squared vibrational frequencies are negative (or equivalently, modes with negative force constants). In general, as the square of the vibrational frequency for a particular structural distortion decreases (or becomes more negative), the greater the tendency of the material towards that structural distortion.

The amplitude, $Q$, of a given mode is defined in the usual way,
\begin{equation}
Q=\sqrt{\sum_i\zeta_{i\alpha}^2},
\end{equation}
where $\zeta_i$ is the displacement of ion $i$ in direction $\alpha$ from its equilibrium position in the parent structure. Our group theoretical analyses were performed with the aid of the ISOTROPY suite of programs\cite{isotropy} and the Bilbao Crystallographic Server.\cite{aroyo06,aroyo06b,aroyo11}

A convenient shorthand notation has also been developed for describing octahedral rotations in perovskites. In Glazer notation,\cite{glazer72} letters represent the magnitude of the tilting distortion about each of the crystallographic axes and a superscript indicates whether the rotations are in-phase or out-of-phase along a particular axis. For example, the rotation pattern in the $R\bar{3}c$ structure is represented as $a^-a^-a^-$ indicating that the rotations are of the same magnitude about each of the $x$, $y$ and $z$ axes and that the rotations are out-of-phase along each axis. The rotation pattern in the $Pnma$ structure is represented as $a^-a^-c^+$: the rotation angles about the $x$ and $y$ axes are the same and out-of-phase, whereas the rotation angle about $z$ is in-phase and different to that of $x$ and $y$. Glazer proposed 23 different `tilt systems', which lead to 15 different space groups. Howard and Stokes subsequently showed using a group-theoretical analysis that there are 15 unique tilt patterns that can occur in real perovskite crystals and these are shown in Table \ref{tilt_systems}.\cite{stokes98} An alternative and equally valid notation for rotations was also developed by Aleksandrov; see Refs. \onlinecite{aleksandrov76} and \onlinecite{aleksandrov01}.  

\begin{table}
\caption{The 15 Glazer tilt systems and corresponding space groups. The tilt systems are listed in decreasing order of their observation in perovskites. The final three tilt systems -- $a^0b^+b^+$, $a^-a^-c^-$, $a^+b^+c^+$ -- have not been observed in bulk materials. Adapted from Ref. \onlinecite{woodward01}.}
\label{tilt_systems}
\begin{tabular}{ll}
Glazer tilt system&Space group\\
\hline
\hline
$a^-b^+a^-$&$Pnma$\\
$a^-a^-a^-$&$R\bar{3}c$\\
$a^+a^+a^+$&$Im\bar{3}$\\
$a^0a^0a^0$&$Pm\bar{3}m$\\
$a^0a^0c^-$&$I4/mcm$\\
$a^0b^-c^+$&$Cmcm$\\
$a^-a^-c^0$&$Imma$\\
$a^0a^0c^+$&$P4/mbm$\\
$a^+b^-c^-$&$P2_1/m$\\
$a^0b^-c^-$&$C2/m$\\
$a^+a^+c^-$&$P4_2/nmc$\\
$a^-b^-c^-$&$P\bar{1}$\\
$a^0b^+b^+$&$I4/mmm$\\
$a^-a^-c^-$&$C2/c$\\
$a^+b^+c^+$&$Immm$\\
\hline
\hline
\end{tabular}
\end{table}

Finally, our first-principles calculations were performed using density functional theory with projector augmented wave potentials, as implemented in VASP. We used the PBEsol functional (which provides an improved description of structural parameters), a 6$\times$6$\times$6 Monkhorst-Pack mesh and a 600 eV plane wave cutoff for all of our calculations. A force convergence tolerance of 2.5 meV/\AA~was used for structural relaxations. Lattice dynamical properties were calculated using density functional perturbation theory. Note that we use the term ``ferroelectric'' to refer to polar structures in which the polarization can in principle be switched to a symmetry-equivalent state with an applied electric field and that satisfy the simple structural criteria devised by Abrahams and co-workers.\cite{abrahams68}

\section*{Competing Instabilities?}
\begin{figure}
\centering
\includegraphics[width=9cm]{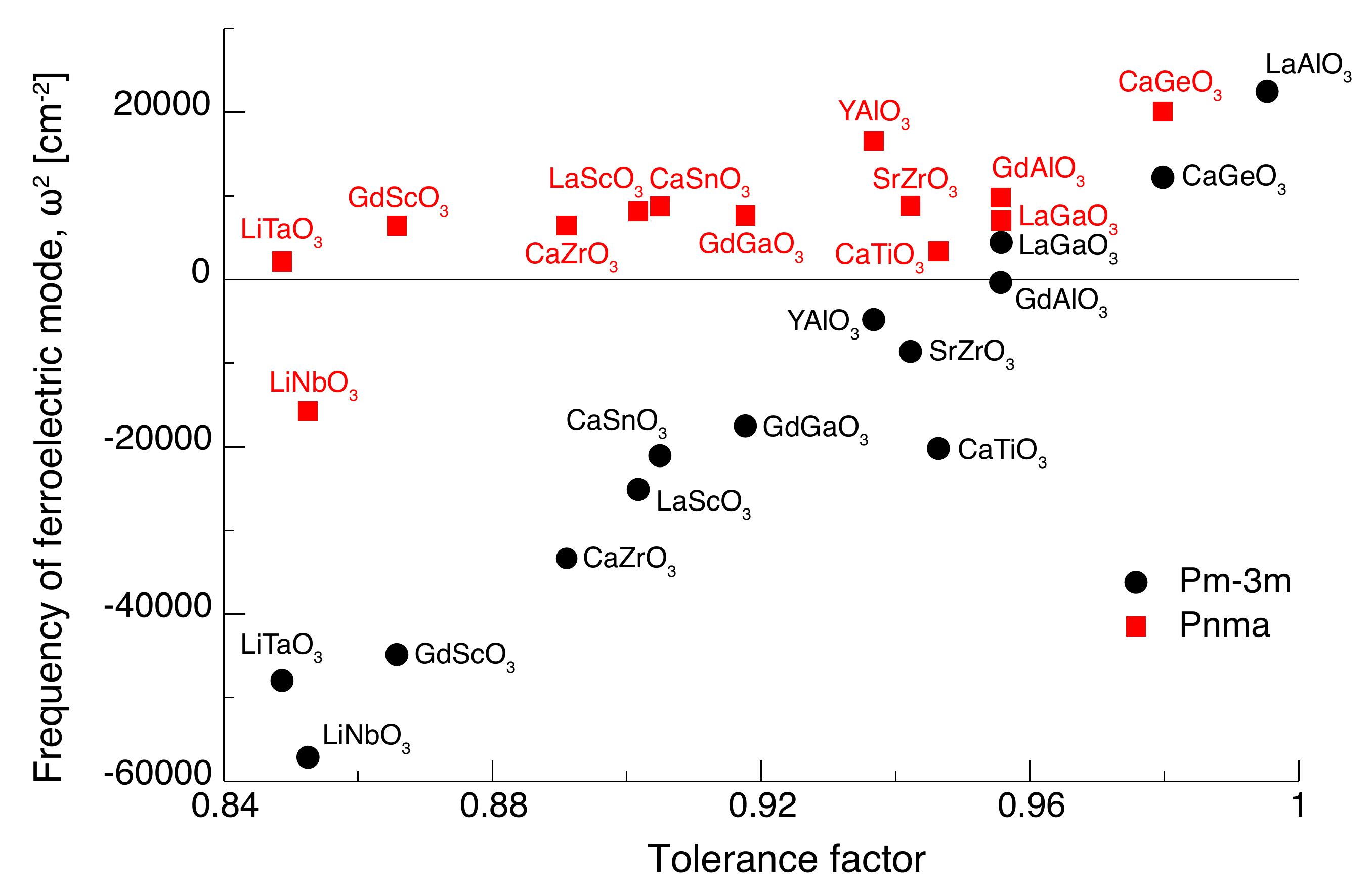}
\caption{\label{Pm-3m_R-3c} Lowest ferroelectric mode frequencies (calculated using density functional perturbation theory) for ABO$_3$ materials in the $Pm\bar{3}m$ and $Pnma$ space groups as a function of bond valence tolerance factor. All of these materials have $Pnma$ ground states, except for LiNbO$_3$ and LiTaO$_3$ (both $R3c$) and LaAlO$_3$ ($R\bar{3}c$).}
\end{figure}

Over a decade of first-principles calculations has taught us that the cubic phases of $Pnma$ perovskites are unstable to both ferroelectric and octahedral rotation distortions.\cite{vanderbilt95,ghosez99,rabe07} Figure \ref{catio3_phonons} shows the phonon dispersion curve of the mineral perovskite, CaTiO$_3$, in the cubic $Pm\bar{3}m$ structure from first-principles calculations. CaTiO$_3$ adopts the $Pnma$ structure in its ground state, which can be obtained by `freezing-in' a combination of the rotation instabilities at the Brillouin zone boundary M and R points. However, Figure \ref{catio3_phonons} shows that the cubic phase is also unstable to a ferroelectric distortion at the zone center although this instability does not persist in the $Pnma$ structure, \textit{i.e.} all the polar zone-center modes are stable in the $Pnma$ structure, suggesting that the rotations have suppressed ferroelectricity. Hence, since most perovskites adopt structures with octahedral rotations, it follows that there are very few perovskite ferroelectrics. The tolerance factor ($t$), a geometric measure of ionic size (mis)match, has been empirically shown to relate this competition to a given ABO$_3$ material's chemical composition:
\begin{equation}
t = \frac{R_{\mathrm{A-O}}}{\sqrt{2}(R_{\mathrm{B-O}})},
\label{t_factor}
\end{equation}
where $R_{\mathrm{A-O}}$ and $R_{\mathrm{B-O}}$ are the ideal A-O and B-O bond lengths for a particular ABO$_3$ material calculated using the bond valence model.\cite{brown78,woodward01} 
In perovskites with $t < 1$, rotations suppress ferroelectricity whereas the opposite occurs for $t\sim 1$. 
The frequency of the lowest ferroelectric mode (from first-principles calculations) for a number of ABO$_3$ materials in both the $Pnma$ and cubic $Pm\bar{3}m$ space groups is plotted in Figure \ref{Pm-3m_R-3c} as a function of each material's tolerance factor. It is evident from Figure \ref{Pm-3m_R-3c} that the squared frequency of the ferroelectric mode for a given material is always higher (less negative) in $Pnma$ than in $Pm\bar{3}m$, which supports the assumption that rotations compete with ferroelectricity. However, the materials with the greatest tendency towards octahedral rotations (those with the smallest tolerance factors) also have the largest ferroelectric instabilities. The latter conclusion is unexpected because prototypical ferroelectrics, such as BaTiO$_3$ and PbTiO$_3$, generally have $t\sim 1$. In these materials the ferroelectric distortion helps to optimize the coordination environment of the B-site cation, which is too small to bond effectively to its neighbors in the cubic phase. In contrast, octahedral rotations help to optimize the coordination environment of the A-site cation for materials with $t < 1$. Implicit in these explanations is the assumption that ferroelectric distortions in ABO$_3$ perovskites can only optimize the B-site, not the A-site, coordination environment. As we shall see however, most perovskites do not display this same B-site optimizing ferroelectric mechanism. An additional question raised by Figure \ref{Pm-3m_R-3c} is why is there only a weak dependence of the ferroelectric mode frequency on tolerance factor for materials in the $Pnma$ structure, whereas there is an almost linear relationship between tolerance factor and mode frequency for materials in $Pm\bar{3}m$? 

\begin{figure}
\centering
\includegraphics[width=8cm]{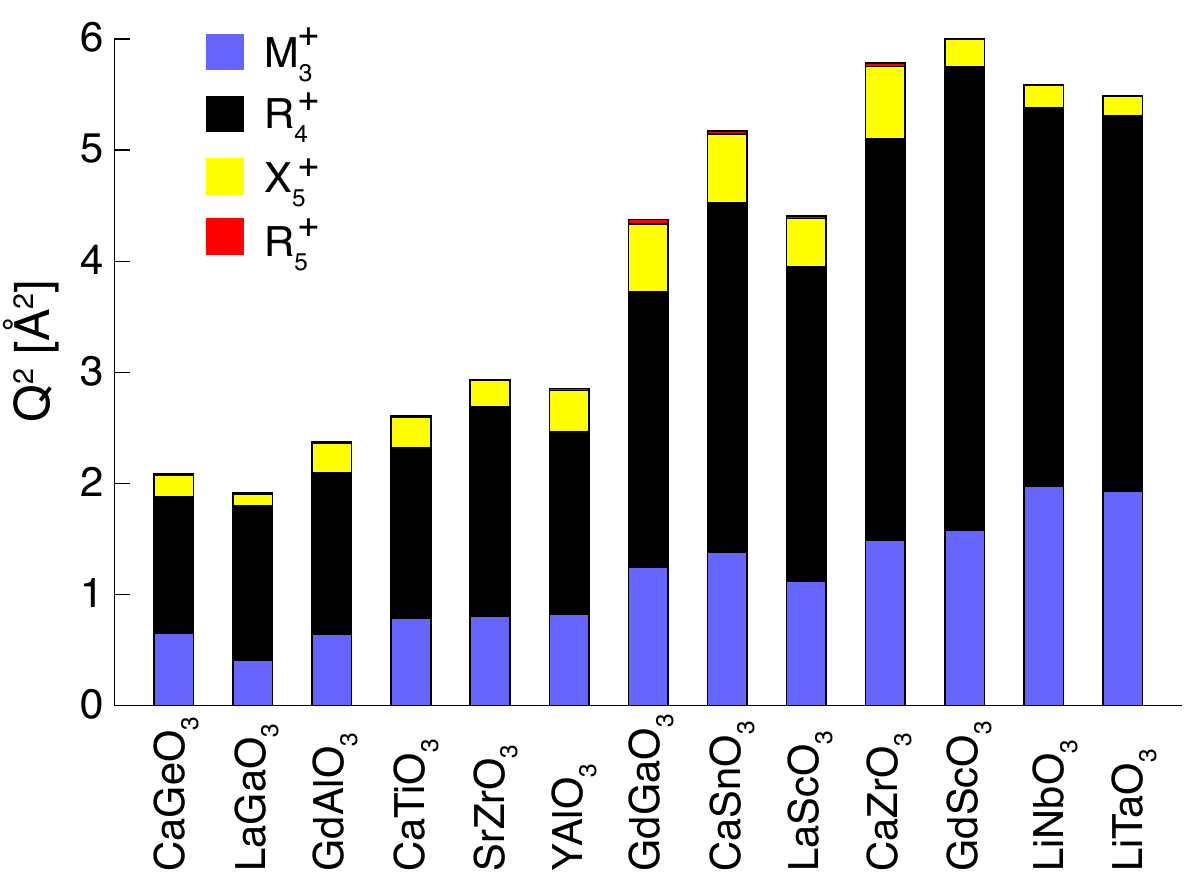}
\caption{Amplitudes of the modes in the fully relaxed $Pnma$ structures of our ABO$_3$ materials. Tolerance factor decreases from left to right.}
\label{mode_decomp}
\end{figure}

\section*{The connection between A-site displacements and the suppression of ferroelectricity in the $Pnma$ structure}
\begin{figure}
\centering
\includegraphics[width=8cm]{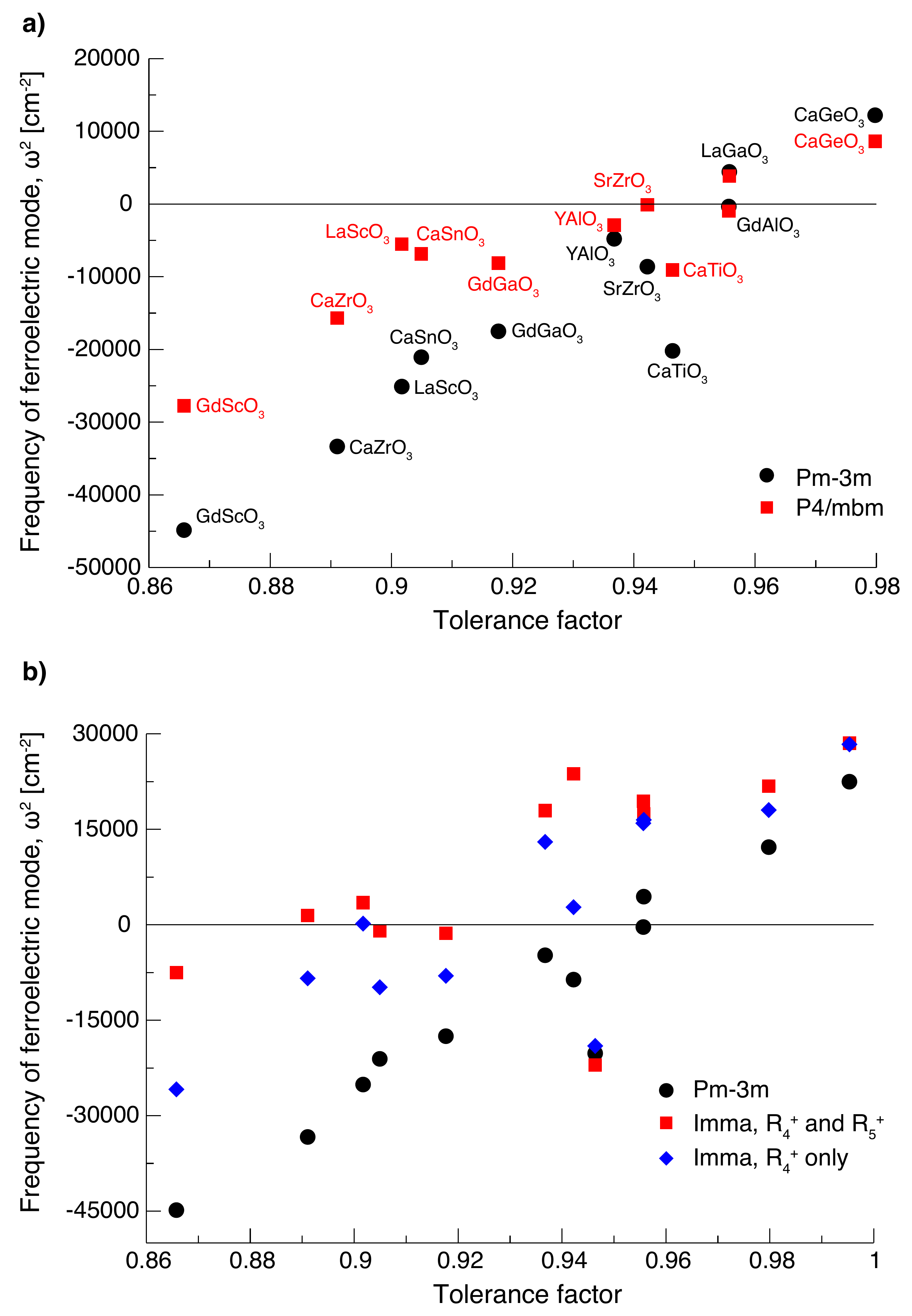}
\caption{\textbf{a}) Lowest ferroelectric mode frequencies for our ABO$_3$ materials in the $Pm\bar{3}m$ and $P4/mbm$ space groups as a function of bond valence tolerance factor. \textbf{b}) As in a) but for the $Imma$ space group. Individual material labels have been omitted for clarity.}
\label{p4mbm_imma}
\end{figure}

\begin{figure*}
\centering
\includegraphics[width=13cm]{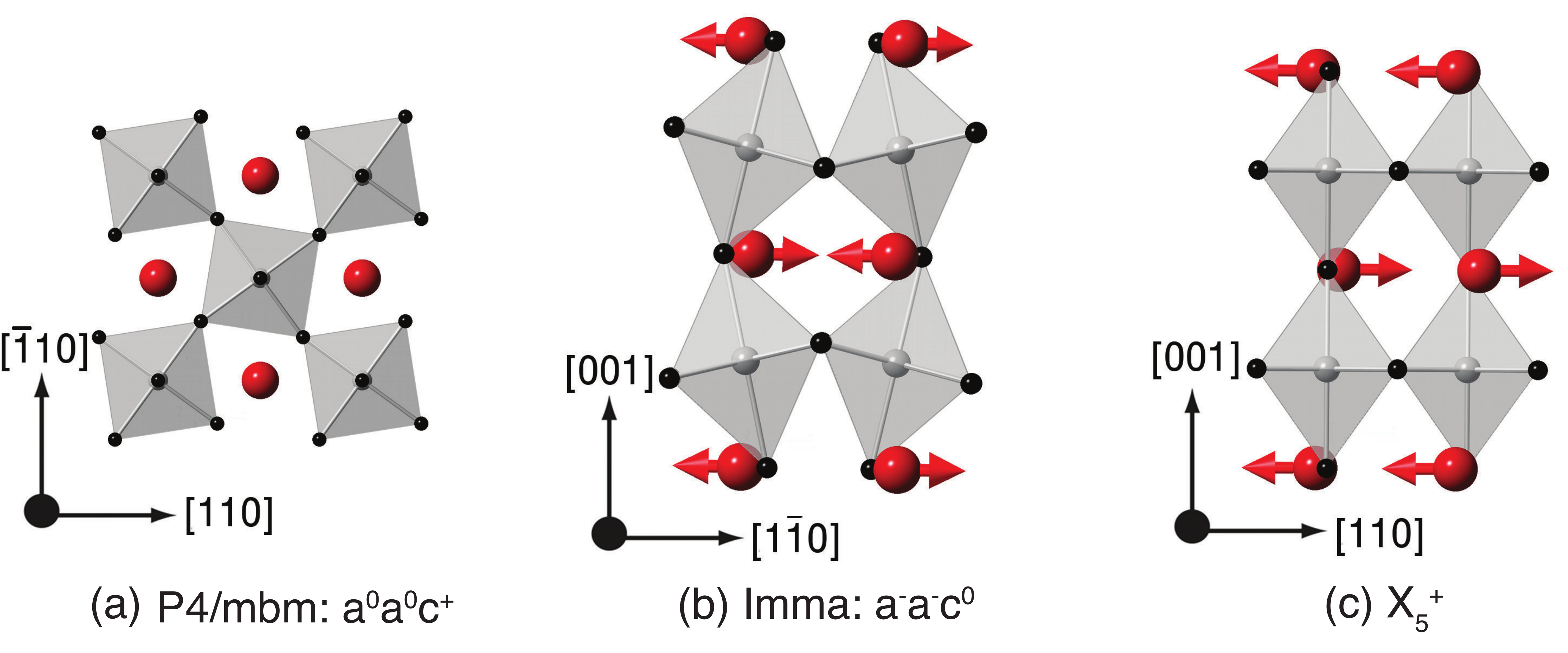}
\caption{The $Pnma$ structure can be described primarily by three different modes of $Pm\bar{3}m$. \textbf{a}) An octahedral rotation about [001], transforming like the irrep $M_3^+$ and leading to space group $P4/mbm$. \textbf{b}) An octahedral rotation about [110], transforming like the irrep $R_4^+$ and leading to space group $Imma$. An A-site displacement mode (transforming like $R_5^+$) accompanies the octahedral rotation but does not change or lower the symmetry further. \textbf{c}) An `antipolar' A-site displacement mode, transforming like the irrep $X_5^+$. This mode becomes energetically favorable only when both $M_3^+$ and $R_4^+$ are present and involves a small displacement of the equatorial oxygen atoms. From Ref. \onlinecite{benedek12}, used with permission.}
\label{pnma_modes}
\end{figure*}

\begin{figure}
\centering
\includegraphics[width=8cm]{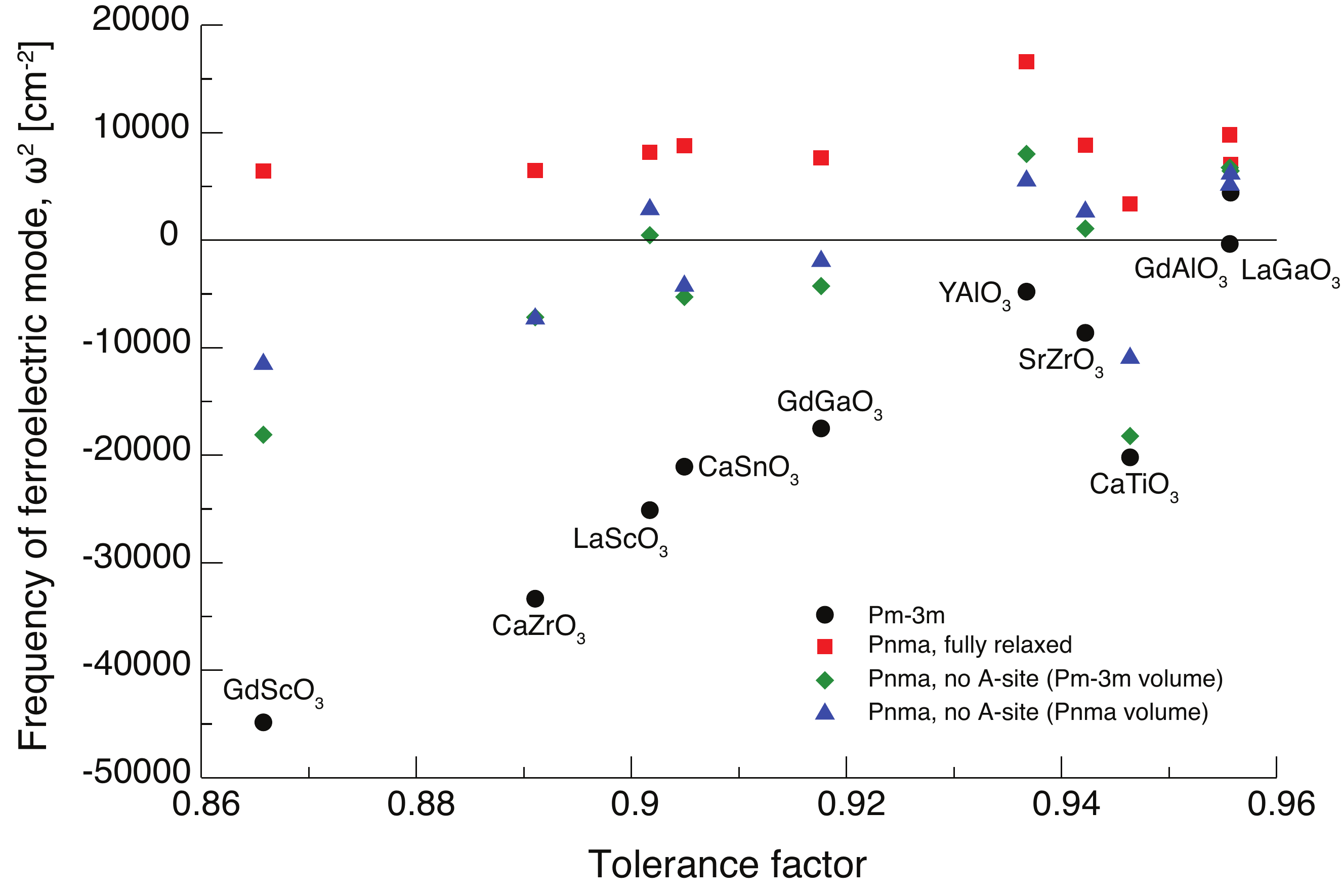}
\caption{Lowest ferroelectric mode frequencies for our ABO$_3$ materials in the $Pm\bar{3}m$ and $Pnma$ space groups.}
\label{fe_pnma}
\end{figure}

The $Pnma$ space group is the most commonly occurring for ABO$_3$ perovskites with octahedral rotations.\cite{woodward01} The symmetry of the $Pnma$ structure is established by a combination of two different rotation patterns, transforming like the irreps $R_4^+$ and $M_3^+$ (the Glazer tilt patterns $a^-a^-c^0$ and $a^0a^0c^+$, respectively; our irrep notation assumes the origin is at the B-site). Although the $Pnma$ structure is most commonly thought of as simply a combination of these two rotations patterns, other kinds of structural distortions are allowed by symmetry. We performed full structural relaxations of the materials shown in Figure \ref{Pm-3m_R-3c} in the $Pnma$ structure and decomposed the relaxed structures into contributions from individual modes using the symmetry-adapted modes of $Pm\bar{3}m$ as a basis. The relaxed structures are in fact built up from contributions from \emph{five} different modes: the two rotation modes and three additional modes ($X_5^+$, $R_5^+$ and $M_2^+$) that are not octahedral rotations and are therefore not accounted for in Glazer notation. Figure \ref{mode_decomp} shows that the rotation modes have the largest amplitudes ($R_4^+$ always has a larger contribution than $M_3^+$), followed by $X_5^+$ and $R_5^+$, both of which involve A-site cation displacements, as shown in Figure \ref{pnma_modes} These modes only become energetically favorable when both $R_4^+$ and $M_3^+$ are present\cite{mulder12} in the case of $X_5^+$ and when $R_4^+$ is present in the case of $R_5^+$. The amplitude of the $M_2^+$ mode is too small to be distinguishable from the other modes and we therefore omit it from further discussion. To investigate the effect of these different distortions on ferroelectricity we performed two sets of computer `experiments'. First, Figure \ref{Pm-3m_R-3c} shows the cumulative effect of all five modes on the frequency of the ferroelectric mode in $Pnma$, but we felt it would be instructive to consider the effect of each mode individually. We performed full structural relaxations of our ABO$_3$ materials in the $P4/mbm$ and $Imma$ space groups, which correspond to the Glazer tilt patterns $a^0a^0c^+$ and $a^-a^-c^0$. We then calculated the frequency of the ferroelectric modes in each space group; the results are shown in Figure \ref{p4mbm_imma}. The $P4/mbm$ structure is purely $M_3^+$ with no contributions from other modes. As Figure \ref{p4mbm_imma}a shows, although this octahedral rotation mode always increases the ferroelectric mode frequency (makes it less negative), the ferroelectric mode is never completely suppressed if it is negative to start with in $Pm\bar{3}m$. The \textit{Imma} structure is built up from contributions from two modes, the rotation mode ($R_4^+$) and an A-site displacement mode ($R_5^+$) as shown in Figure \ref{pnma_modes}. Similar to $P4/mbm$, Figure \ref{p4mbm_imma}b shows that the rotation mode always increases the ferroelectric mode frequency, although generally not enough to completely suppress ferroelectricity. However, when the A-site displacement $R_5^+$ mode is included, the ferroelectric mode is suppressed much more strongly. For the second experiment, we fully relaxed our materials in the $Pnma$ structure, in one case keeping the volume fixed to that of the $Pm\bar{3}m$ cell (the volume was kept constant and equal to $Pm\bar{3}m$ with no orthorhombic strain) and in the other allowing it to fully relax. Then, keeping the rotations at their relaxed values, we `zeroed' the amplitude of the $X_5^+$ mode and the A-site component of $R_5^+$, \textit{i.e.}, we returned the A-site cations to their ideal $Pm\bar{3}m$ positions. Figure \ref{fe_pnma} shows that, again, the rotations always raise the ferroelectric mode frequency but as the tolerance factor decreases (and the ferroelectric instability grows larger), it is the A-site displacements that play the decisive role in suppressing ferroelectricity in $Pnma$ (we discuss this more extensively below). These A-site displacement modes hold the key to understanding the data displayed in Figure \ref{Pm-3m_R-3c} and we now return to the puzzle of why materials with small tolerance factors have large ferroelectric instabilities $Pm\bar{3}m$, which are suppressed in $Pnma$. 

As we have already discussed, ABO$_3$ materials with $t < 1$ have undercoordinated A-sites in $Pm\bar{3}m$ and generally adopt ground state structures with octahedral rotations. However, a ferroelectric distortion from $Pm\bar{3}m$ could also optimize the A-site coordination environment if it involved a significant A-site component. A-site driven ferroelectricity is not a new phenomenon. Perhaps the most well-known example of an A-site driven ferroelectric is Pb(Zr,Ti)O$_3$, although in this material the ferroelectric distortion is driven primarily by the stereochemical activity of the Pb$^{2+}$ lone pair, rather than geometrical factors. We examined the ferroelectric modes for all the materials in our test set in the $Pm\bar{3}m$ structure and found that the contribution of the A-site to the ferroelectric eigenvector was always significantly larger than the B-site contribution (see Table S1 in the Supporting Information). Hence, the smaller an ABO$_3$ material's tolerance factor, the more unfavorable the coordination environment at the A-site and the greater the tendency towards a ferroelectric distortion; this is exactly the trend shown in Figure \ref{Pm-3m_R-3c}. However, a rotation distortion optimizes the A-site more effectively in these materials and so the ferroelectric distortion is suppressed, but as the tolerance factor decreases and the ferroelectric instability in the cubic phase grows larger, \emph{it is the A-site displacements that accompany the rotations in the Pnma structure that play the decisive role in suppressing ferroelectricity.} Our results concerning the connection between A-site displacements and ferroelectricity are also consistent with the observations of Mulder and co-workers, who showed for a series of perovskites that the force constant of the $X_5^+$ mode in $Pm\bar{3}m$ has the same linear dependence on the tolerance factor as the ferroelectric instability.\cite{mulder12} 

Not all octahedral rotation patterns allow the A-site to shift from its ideal $Pm\bar{3}m$ position. As explained by Woodward (and Thomas\cite{thomas96}) for the $Pnma$ structure,\cite{woodward97} the extra degree of freedom afforded by A-site displacements allows the A-site to further optimize its coordination environment by resulting in a better distribution of A-O bond distances. Hence, for ABO$_3$ perovskites in the $Pnma$ space group, the A-site coordination environment is optimized by \emph{both} octahedral rotations and A-site displacements. This explains the weak correlation between the tolerance factor and ferroelectric mode frequency for $Pnma$ shown in Figure \ref{Pm-3m_R-3c}. In $Pnma$, there is no driving force for \emph{further} movement of the A-site, since the bonding preferences have already been satisfied by the octahedral rotations and the A-site displacements that accompany them. 

\begin{figure}
\subfloat[][]{
\centering
\begin{tabular}{lcc}
Material&$Pnma$, fully relaxed&$Pnma$, with $Q_{X_5^+}$ = 0\\
\hline
\hline
YAlO$_3$&-0.13&0.12\\
GdAlO$_3$&-0.08&0.10\\
LaGaO$_3$&-0.02&0.03\\
GdScO$_3$&-0.11&0.34\\
CaZrO$_3$&0.04&0.14\\
CaSnO$_3$&-0.07&0.13\\
LaScO$_3$&-0.11&0.07\\
GdGaO$_3$&-0.22&0.22\\
SrZrO$_3$&-0.19&-0.13\\
CaTiO$_3$&-0.03&0.07\\
\hline
\hline
\end{tabular}}
\\
\subfloat[][]{
\centering
\includegraphics[width=5cm]{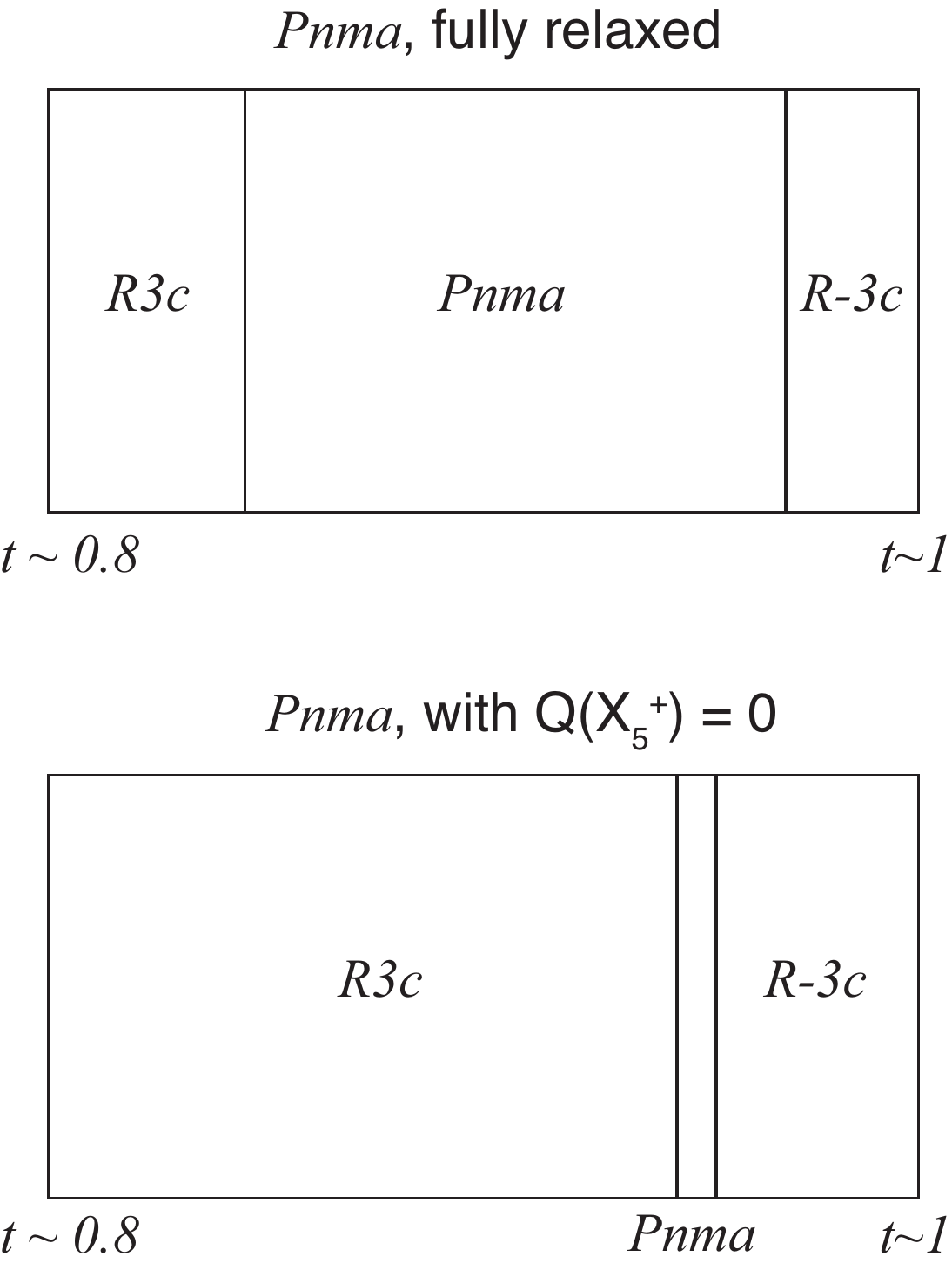}
}
\caption{\label{ediff}a) Energy difference, in eV per formula unit, between $Pnma$ and $R3c$ ($R\bar{3}c$ for YAlO$_3$, GdAlO$_3$ and LaGaO$_3$) for $Pnma$ structures with relaxed ions and for those with anti-polar A-site displacements zeroed out. A negative energy difference indicates that the $Pnma$ structure is more stable than $R3c$/$R\bar{3}c$. b) Schematic of the ABO$_3$ phase diagram as a function of tolerance factor (top) and a hypothetical phase diagram in which anti-polar A-site displacements have been suppressed in the $Pnma$ structure (bottom).} 
\end{figure}

To further explore the role and extent to which the A-site displacements stabilize the $Pnma$ structure, we performed an additional set of calculations where we took materials that display the $a^-a^-c^+$ ($Pnma$) rotation pattern in their ground states and asked the question what would their properties be if they displayed the $a^-a^-a^-$ ($R\bar{3}c$) rotation pattern (note that in contrast to $Pnma$, the $R\bar{3}c$ structure involves only the $R_4^+$ irrep, rather than both $R_4^+$ and $M_3^+$). Most of these $R\bar{3}c$ structures display ferroelectric instabilities  (which indicates that a ferroelectric distortion to the polar group $R3c$ is energetically favorable; $R3c$ is a low energy metastable phase for the materials we considered). 
The main cation contribution to the ferroelectric mode for all the materials we investigated comes from the A-site, as shown in the Supporting Information. We fully relaxed the ions for both materials in the $Pnma$ and $R3c$ structures (or $R\bar{3}c$ for materials that do not have ferroelectric instabilities in this space group) keeping the volume fixed to that of $Pm\bar{3}m$ to facilitate comparison. Then as before, keeping the rotations fixed at their relaxed values in the $Pnma$ structure, we zeroed the $X_5^+$ mode, the A-site displacement mode with the largest amplitude, as shown in Figure \ref{mode_decomp}. Figure \ref{ediff}a shows that suppressing the A-site displacements of the $X_5^+$ mode (the column labelled $Q_{X_5^+} = 0$) destabilizes the $Pnma$ structure such that it is no longer the lowest energy (with the exception of SrZrO$_3$, however the energy of $Pnma$ is still raised). These calculations illustrate the very significant stabilizing effect of the A-site displacements in the $Pnma$ structure. Not only do they suppress ferroelectricity, but they also contribute crucially to the stability of this structure. Woodward\cite{woodward97} performed a similar set of computer experiments on YAlO$_3$, using a classical shell model potential to calculate and compare the energy of YAlO$_3$ in the $R\bar{3}c$ and $Pnma$ structures. Like us, he found that the energies of the two polymorphs were similar when A-site displacements were suppressed. Our results, together with those of Woodward and Thomas, leave little doubt as to the role and importance of A-site displacements in stabilizing the $Pnma$ structure.

To highlight the significance of this result we present a schematic phase diagram in Figure \ref{ediff}b, which shows  the sequence of structures adopted by perovskites as a function of tolerance factor. At high tolerance factors there is a narrow range of $R\bar{3}c$ materials, followed by a wide range of  $Pnma$ structures (most perovskites with tolerance factors less than one adopt this structure), and finally those with very small tolerance factors ($t < 0.9$) tend to be $R3c$, the polar subgroup of $R\bar{3}c$ (this sequence of structures can be understood by noting that the A-site coordination environment changes from 9 in $R\bar{3}c$ to 8 in $Pnma$ and 6 in $R3c$; this is discussed in greater detail in the Supporting Information). If the anti-polar A-site displacements were suppressed however, most $Pnma$ compounds would be metastable and the $R3c$ (and to some extent, $R\bar{3}c$) structure would have a much wider stability window. We emphasize that Figure \ref{ediff}b is intended merely as a schematic to highlight the effect of A-site displacements on the ABO$_3$ phase diagram. That is, we are not suggesting that suppressing A-site displacements in the $Pnma$ structure is a practical route to obtaining new perovskite ferroelectrics. We discuss synthesis possibilities in greater detail in the next section.

Before answering the question posed in the title of this manuscript, we pause to discuss in general, phenomenological terms the interaction between ferroelectricity and octahedral rotations in ABO$_3$ materials (the generalization to explicitly consider $Pnma$ is trivial and none of the arguments we are about to make depend on it). One can quantify the competition between the distortions by expanding the free energy ($\mathcal{F}$) about the cubic $Pm\bar{3}m$ phase to obtain (to fourth order):
\begin{equation}
\mathcal{F}=\alpha_{02} Q^2_{FE} + \alpha_{20}Q^2_R + \beta_{02} Q^4_{FE} + \beta_{20}Q^4_R + \beta_{22}Q^2_{FE}Q^2_R,
\label{free_energy}
\end{equation}  
where $Q_{FE}$ and $Q_R$ are the amplitudes of the ferroelectric and octahedral rotation modes and the Greek letters are coefficients (note that $Q_R$ refers generally to the amplitude of any rotation mode). To lowest order ferroelectricity and octahedral rotations are coupled through the biquadratic term, $\beta_{22}$. From (\ref{free_energy}), the squared frequency of the ferroelectric mode in $Pm\bar{3}m$ ($\omega^2_{FE}$) can be written as,
\begin{equation}
\omega^2_{FE} \sim \alpha_{02} + \beta_{22}Q^2_R.
\end{equation}
$\alpha_{02}$ is negative for materials that have a ferroelectric instability in $Pm\bar{3}m$. If $\beta_{22}$ is negative, $\omega^2_{FE}$ decreases (becomes more negative) and therefore ferroelectricity is favored by the presence of octahedral rotations. If $\beta_{22}$ is positive, $\omega^2_{FE}$ will increase, indicating that ferroelectricity is suppressed by octahedral rotations. For all of the materials that we have studied so far, rotations (regardless of the tilt pattern) raise the ferroelectric mode frequency and hence $\beta_{22}$ is positive. For materials with larger tolerance factors however, the ferroelectric energy scale, $\omega^2_{FE}$, is much less negative than the positive contribution coming from the coupling to rotations, $\beta_{22}R^2$, and therefore one should not really speak of a competition between rotations and ferroelectricity (though perhaps this is more of a semantic point\cite{note_energies}). It is in this regime (materials for which the ferroelectric instability is not so large in $Pm\bar{3}m$, such as SrZrO$_3$\cite{amisi12}) that rotations are enough to completely suppress ferroelectricity and A-site displacements merely act to further harden the polar mode. As the tolerance factor decreases, $\alpha_{02}$ becomes increasingly negative and rotations and ferroelectricity do in fact start competing with each other. However, for the range of tolerance factors in which this competition occurs, we have shown for the first time that ferroelectricity wins, but is nonetheless suppressed by the A-site displacements. For example, for materials such as GdScO$_3$ and CaZrO$_3$, rotations by themselves do not suppress ferroelectricity. Hence, we believe that the answer to the question of whether rotations suppress ferroelectricity really is `it depends': for larger tolerance factor materials with rotations it is the rotations alone that suppress ferroelectricity (but here we shouldn't speak of a competition), but for many $Pnma$ materials with smaller tolerance factors this is not the case, \textit{i.e.} A-site displacements are needed to completely suppress ferroelectricity.

Why are there so few perovskite ferroelectrics? Because most perovskites adopt the $Pnma$ structure in which A-site displacements suppress ferroelectricity. Although we think of prototypical ferroelectrics as being B-site driven with $t > 1$, there are relatively few such compounds; that is, there are very perovskites with $t > 1$ and of those only a very small number have ferroelectrically active cations (such as Ti) on the B-site. As our results demonstrate, most perovskites would be A-site driven geometric ferroelectrics, if their ferroelectric phases could be stabilized. Why then do so many perovskites adopt the $Pnma$ structure? As explained by Woodward,\cite{woodward97} the $a^-a^-c^+$ tilt pattern maximizes A-O covalent bonding interactions and minimizes A-O repulsion. The A-site displacments in $Pnma$ play an important role in minimizing A-O repulsion and optimizing the A-site coordination environment, as also supported by our first-principles calculations.

Is there any way to circumvent the suppressing effect of the rotations/A-site displacements or to promote the coexistence of rotations and ferroelectricity? One approach is to substitute the A-site cation with a lone pair active cation. In oxides containing lone pair cations, the cation n$s$-O 2$p$ interaction results in a set of bonding and antibonding states, both of which are filled.\cite{payne06,stoltzfus07,walsh11} The occupied n$s$-O 2$p$ antibonding levels are stabilized by mixing with the cation $p$ states, which reside in the conduction band. However, since $s$ and $p$ states are of different parity the cation site symmetry must be such that $sp$ mixing is allowed, in which case a second-order Jahn-Teller distortion to a lower symmetry -- often polar -- structure is favored. For example, LaFeO$_3$ adopts the nonpolar $Pnma$ structure. If the A-site is replaced by Bi$^{3+}$, which has a lone electron pair and is approximately the same size as La$^{3+}$, then we obtain the polar $R3c$ structure instead, \textit{i.e.}, the lowest energy structure changes from $Pnma$ for LaFeO$_3$ to $R3c$ for BiFeO$_3$. However, only a small group of elements have lone electron pairs and in any case, the presence of a lone pair cation does not guarantee a polar structure, \textit{e.g.}, BiScO$_3$\cite{belik06b} and BiGaO$_3$\cite{belik06} adopt nonpolar structures, despite the lone pair on Bi.

Another possible route to inducing ferroelectricity in nominally non-ferroelectric materials is to grow the material as an epitaxial thin-film under biaxial strain. SrTiO$_3$, which is nominally not ferroelectric (it is a quantum paraelectric), can be made ferroelectric at room temperature when grown under tensile strain.\cite{schlom04} A polar or ferroelectric distortion optimizes the B-site coordination environment, which is destabilized by the increase in the in-plane lattice parameter. G\"{u}nter and co-workers also recently showed that ferroelectricity can be induced by strain in the $Pnma$ perovskite CaMnO$_3$.\cite{gunter12}  

Our results suggest an alternative route to the creation of new ferroelectrics. Select a material with a small tolerance factor and stabilize it in a rotation pattern that forbids A-site displacements, such that any further optimization of the A-site coordination environment must come from a ferroelectric distortion. As we shall see in the next section, a number of these materials already exist.

\section*{Ilmenite-derived $R3c$ materials as a family of A-site driven geometric ferroelectrics}
\begin{figure}
\centering
\includegraphics[width=8cm]{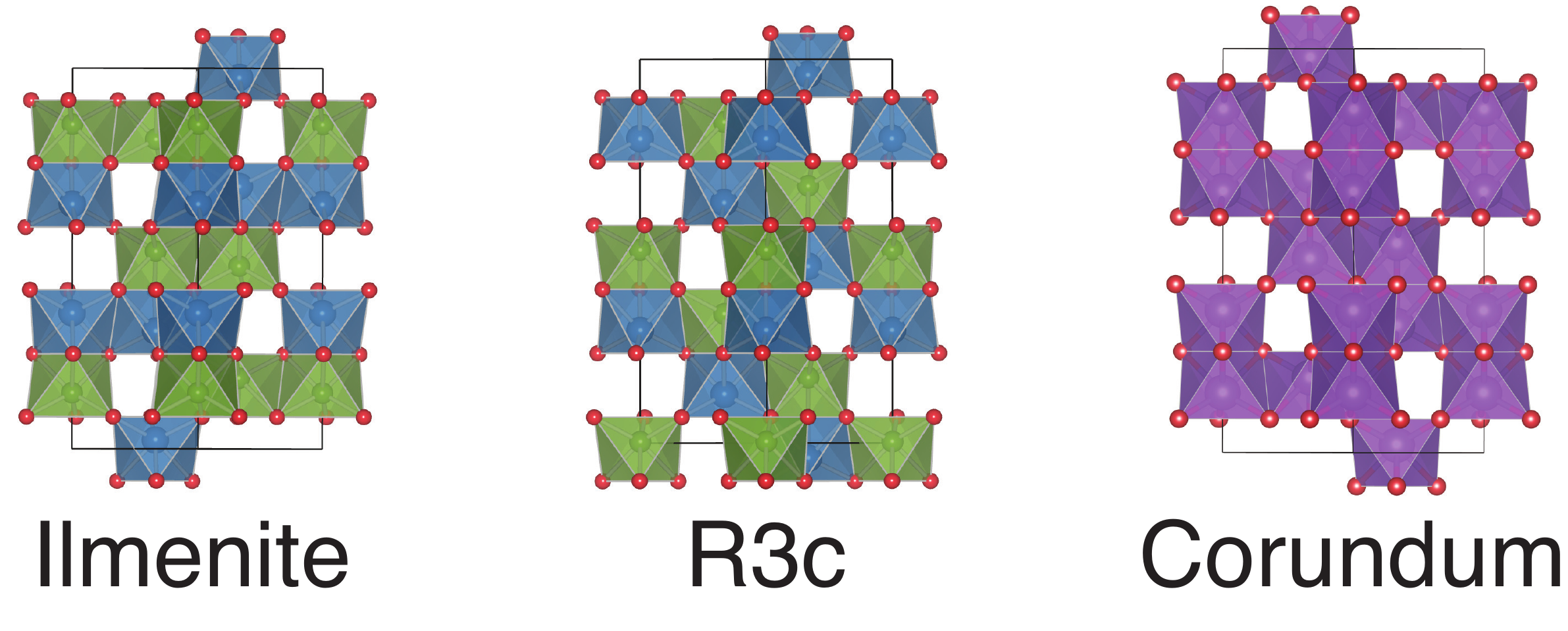}
\caption{\label{ilmenite_r3c_corundum}The Ilmenite, $R3c$ (LiNbO$_3$) and Corundum structure types shown in the hexagonal setting. The ilmenite and $R3c$ structures can be understood as ordered variants of corundum.\cite{navrotsky98} In the ilmenite structure, the A and B-site cations order in alternate layers whereas in $R3c$, each layer contains equal proportions of A and B-site cations. The coordination polyhedra in the corundum structure are a different color to those in the Ilmenite and $R3c$ images to indicate that all cation sites are crystallographically equivalent.}
\end{figure}
Using a combination of group theoretical arguments and first-principles calculations, it was recently shown that the mineral Ilmenite, FeTiO$_3$, is a multiferroic with strong polarization-magnetization coupling when it is stabilized in $R3c$.\cite{fennie08,varga09} A number of other oxides form in the Ilmenite structure (space group $R\bar{3}$, $\#$148) but can be converted to the perovskite ($Pnma$) or $R3c$ structure under high-pressure--high-temperature conditions.\cite{sleight70,goodenough73,navrotsky98,mizoguchi04} Both the A and B-site are octahedrally coordinated in the ilmenite and $R3c$ structures and both structure types can be considered ordered variants of the corundum structure (see Figure \ref{ilmenite_r3c_corundum}). We speculate in the Supplementary Information as to the structural features which lead particular materials to adopt either the ilmenite, $R3c$ or $Pnma$ structures. 

\begin{figure}
\centering
\includegraphics[width=7.5cm]{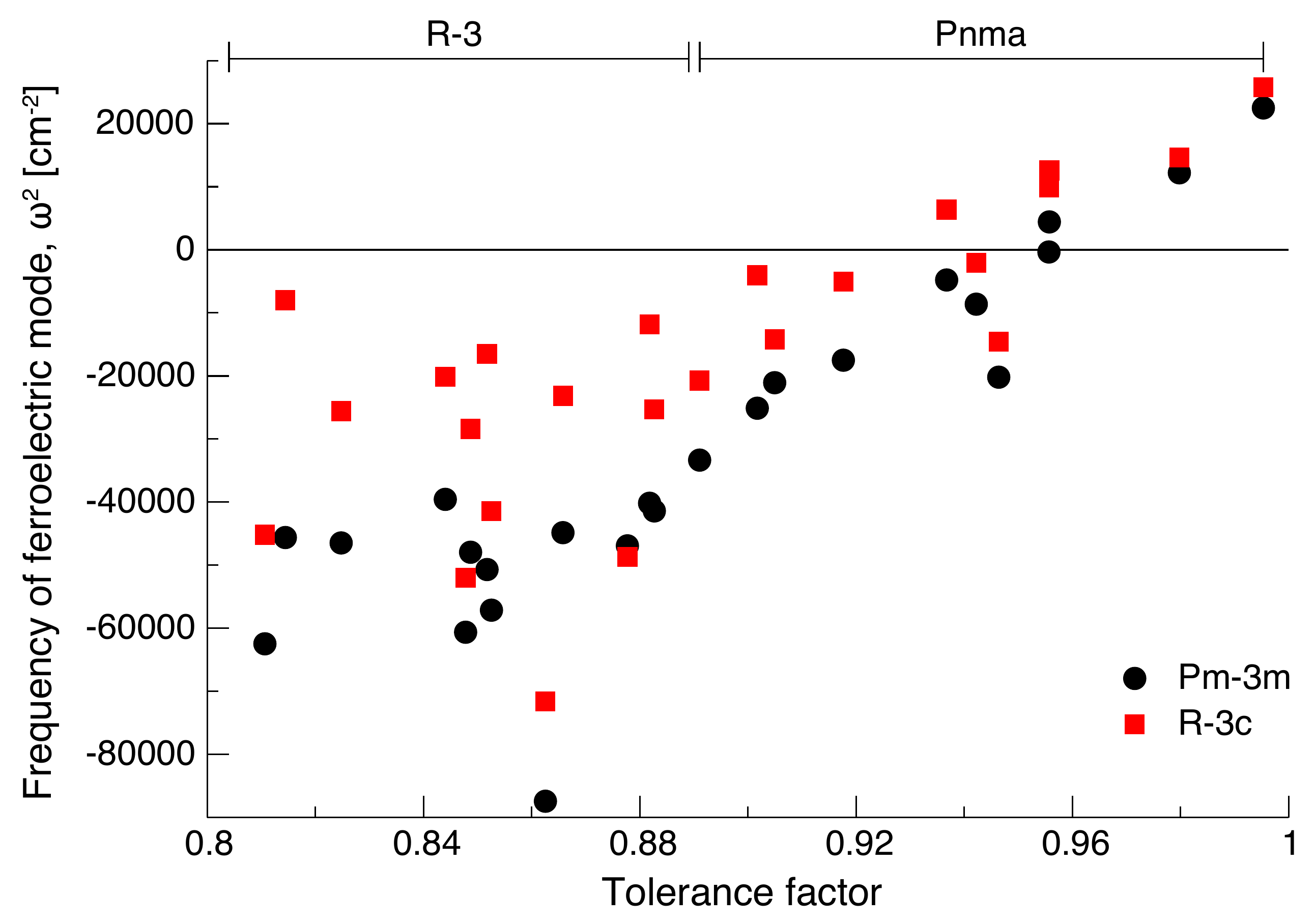}
\caption{Frequency squared of ferroelectric mode in $Pm\bar{3}m$ and $R\bar{3}c$ versus tolerance factor for the materials we investigated. For the sake of clarity, we have omitted labels for the data points. This figure can be found in table form in the Supporting Information. The $R\bar{3}$ `region' consists mostly of materials that form in the Ilmenite structure whereas the \textit{Pnma} region consists mostly of materials that form in the orthorhombic perovskite structure (this is the same $Pnma$ data that is depicted in Figure \ref{Pm-3m_R-3c}).}
\label{ilmenites}
\end{figure}

\begin{figure}
\centering
\includegraphics[width=6cm]{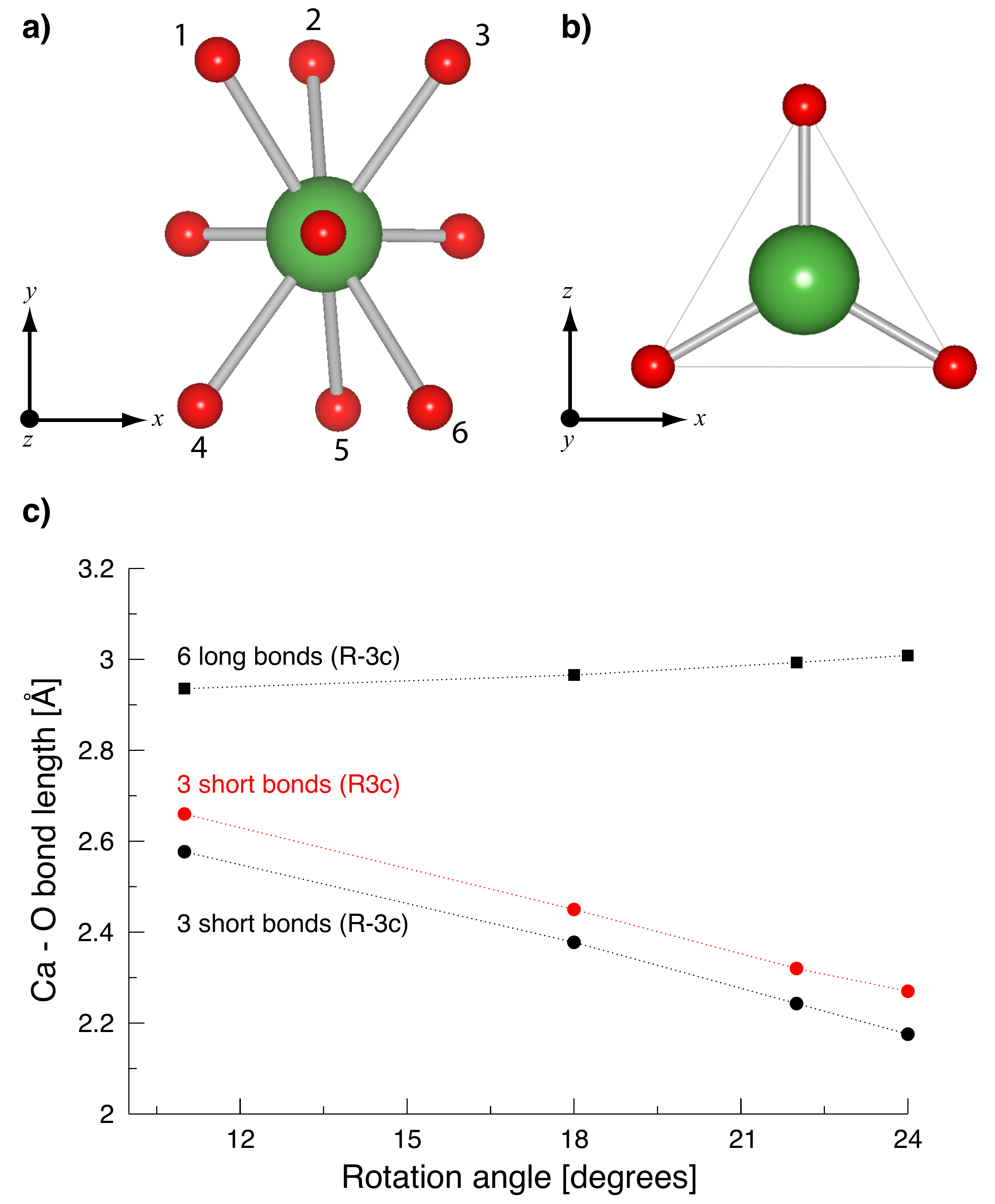}
\caption{\label{a-site_coord}A-site coordination environment in $R\bar{3}c$ space group. The large (green) sphere is the A-site cation whereas the small (red) spheres are the oxygen ions. The A-site is nine coordinate with six longer bonds and three shorter bonds to oxygen. \textbf{a)} The oxygen ions labeled 1-6 form the six long bonds. \textbf{b)} The unlabeled ions in \textbf{a)} form the three short bonds. \textbf{c)} Comparison of bond length changes (at constant volume) in $R\bar{3}c$ and $R3c$ CaZrO$_3$ as a function of rotation angle in $R\bar{3}c$. We took CaZrO$_3$ as a representative example, but obtained qualitatively the same results for other ABO$_3$ materials in our test set. The lines are guides to the eye.}
\end{figure}

The search for multiferroic and lead-free ferroelectric and piezoelectric materials has lead to renewed interest in the $R3c$/ilmenite family of materials. For example, ZnSnO$_3$ was recently synthesized in the $R3c$ structure both through a high-pressure route\cite{inaguma08} and in thin-film form (ferroelectric switching was also demonstrated\cite{son09}). Aimi and co-workers used high-pressure techniques to synthesize $R3c$ MnTiO$_3$ and MnSnO$_3$ and observed spin-canting at low temperatures.\cite{aimi11} The presence of ZnSnO$_3$ and MnSnO$_3$ in the list of materials just discussed is surprising, since neither contains a lone-pair cation (Sn is in a 4+ oxidation state) or otherwise ferroelectrically active ($d^0$) cation. We calculated the relaxed structures and ferroelectric mode frequencies for a number of Ilmenite-structured materials in both the $Pm\bar{3}m$ and $R\bar{3}c$ space groups (here we are taking $R\bar{3}c$ as our non-polar paraelectric parent structure; $R\bar{3}c$ can be related to $R3c$ by a ferroelectric mode that transforms like the irrep $\Gamma_2^-$). Figure \ref{ilmenites} shows that there is a strong correlation between tolerance factor and ferroelectric mode frequency in both the $Pm\bar{3}m$ and $R\bar{3}c$ space groups for all the materials we studied. In contrast to $Pnma$, the rotation pattern that produces the $R\bar{3}c$ structure transforms like the single irrep $R_4^+$ ($a^-a^-a^-$). A-site cation displacements are forbidden by symmetry in this space group 
and so octahedral rotations are the only mechanism by which the A-site can optimize its coordination environment. However, the magnitude of the rotation distortion is limited by the anion-anion contacts, which become increasingly shorter as the rotation angle increases (the three shortest A-O bonds also become increasingly shorter, as Figure \ref{a-site_coord}c shows). For materials with very small tolerance factors ($t \sim< 0.95$), the rotation distortion in $R\bar{3}c$ becomes electrostatically (and energetically) unfavorable before an optimal bonding environment is established for the A-site. A ferroelectric distortion to $R3c$ can further optimize the bonding environment at the A-site if it involves a significant movement of the A-site cation. The ferroelectric mode that takes the $R\bar{3}c$ structure to $R3c$ involves significant displacement of the A-site cation (see Supporting Information) for all the ABO$_3$ materials we studied -- \emph{including the Ilmenites} -- such that the three shortest A-O bonds are lengthened, \textit{i.e.} the bonds shown in Figure \ref{a-site_coord}b. These three oxygens, along with those numbered 1-3 in Figure \ref{a-site_coord}a, form the octahedral coordination environment in $R3c$ (Table S4 in the Supporting Information compares the A-O bond lengths in the $R\bar{3}c$ and $R3c$ structures for all the materials we considered; the three shortest A-O bonds are always longer in $R\bar{3}c$ than $R3c$). The driving force for the distortion can also be understood from bond valence arguments. Although the A-site is technically nine coordinate in $R\bar{3}c$, the majority of the bonding is confined to the three oxygens that form a coplanar environment around the A-site (Figure \ref{a-site_coord}), which reduces the effective coordination number to three. For example, for CaZrO$_3$ the bond valence of the three short bonds is 0.349 whereas that of the six long bonds is just 0.067. However in the $R3c$ structure, six oxygens effectively coordinate the A-site. Again, using CaZrO$_3$ as an example, the bond valence of the three shorter Ca-O bonds is 0.290 and the bond valence of the three longer Ca-O bonds is 0.243. Essentially what this analysis is telling us is that in the $R3c$ structure, the A-site coordination number is maximized and hence so is the number of orbitals on the A-site that can participate in bonding.\cite{woodward97}

The A-site displacement that takes the $R\bar{3}c$ structure to $R3c$ is analogous to that which accompanies the octahedral rotations in $Pnma$. There is a strong correlation between the tolerance factor and ferroelectric mode frequency for $Pm\bar{3}m$ and $R\bar{3}$c because in these space groups, the A-site displacements of the ferroelectric mode play a significant role in optimizing the bonding environment. Also notice in Figure \ref{p4mbm_imma} that there is a reasonable dependence of the ferroelectric mode frequency on tolerance factor in the $Imma$ space group (considering the rotation mode only) and in $P4/mbm$, a space group in which A-site displacements are symmetry-forbidden. Of the materials we investigated, FeSnO$_3$ and MgSnO$_3$ appear particularly promising as new $R3c$ ferroelectrics. The energy difference between the Ilmenite and $R3c$ structure is only 5 meV for MgSnO$_3$; our calculations predict that $R3c$ FeSnO$_3$ is actually \emph{more stable} than the Ilmenite form and is therefore a particularly exciting candidate. In addition, FeSnO$_3$ satisfies exactly the same symmetry criteria as FeTiO$_3$ and is therefore also a strongly coupled multiferroic. As far we are aware, it has not yet been synthesized.

We have identified the structural features that may make a particular material a good candidate for a new ferroelectric or polar functional material, \textit{e.g.,} small tolerance factor, large octahedral rotation distortions. Such materials are likely to be metastable phases and hence the synthesis challenge lies in devising a procedure to stabilize the ferroelectric structure. For the Ilmenites, high-pressure, high-temperature synthesis methods provide an established and reliable route to obtain the $R3c$ phase and recent work (discussed above\cite{son09}) has shown that the $R3c$ phase can also be obtained in thin-film form for some materials. In addition, the results presented in Figure \ref{ediff}a suggest that for some materials, the $Pnma$ and $R3c$ phases are reasonably close in energy and that in these cases, it may be possible to stabilize (using thin-film growth techniques, for example) the $R3c$ phase over $Pnma$. 

\section*{Conclusions}
We have investigated the interaction between ferroelectricity and octahedral rotations in a group of ABO$_3$ perovskite oxides and have revealed the key role of the coordination preferences of the A-site cation in suppressing ferroelectricity in ABO$_3$ materials with small tolerance factors and in stabilizing the $Pnma$ structure, the most common space group adopted by ABO$_3$ perovskites. Although materials such as BaTiO$_3$ and PbTiO$_3$ are considered prototypical perovskite ferroelectrics, the majority of perovskites will not display the same ferroelectric mechanism (B-site off-centering) as these materials. Ferroelectric distortions in perovskites with $t < 1$ are driven by A-site coordination preferences and do not generally involve charge transfer.\cite{ederer06} Recent work by us on layered and double perovskites has revealed the existence of yet another ferroelectric mechanism that also involves A-site coordination preferences.\cite{benedek11,benedek12,mulder12}. Hence, A-site displacements play a much bigger role in the ferroelectric mechanisms of ABO$_3$ materials than was previously appreciated. 

Our results also clarify the origin of ferroelectricity in Ilmenite-derived $R3c$ materials, such as the electric-field controllable weak ferromagnet FeTiO$_3$. When FeTiO$_3$ was first discussed in the context of multiferroics design,\cite{fennie08} it was implied that a ferroelectrically active $d^0$ cation was required on the B-site for the ferroelectric part of the mechanism. In fact, even though the B-site does contribute to the ferroelectricity in FeTiO$_3$, the larger contribution comes from the A-site. This helps to explain why other materials in the same family, such as ZnSnO$_3$, are ferroelectric even though they do not contain any ferroelectrically-active cations. This family of materials also illustrates the counter-intuitive design principle established in light of learning the role of A-site displacements in suppressing ferroelectricity in and stabilizing the $Pnma$ structure: Ilmenite-derived $R3c$ materials that have very small tolerance factors, large rotations and a nonpolar parent structure -- $R\bar{3}c$ -- in a space group in which A-site displacements are forbidden by symmetry. 

The functional properties of perovskites were discovered over 50 years ago and have been the subject of intense study ever since. Our work sheds new light on the chemical and physical factors that drive structural distortions in ABO$_3$ materials and suggests that there are still surprises to be discovered in this fascinating class of materials, particularly for materials with low tolerance factors, for example, those at the  boundary of Ilmenite-LiNbO$_3$-$Pnma$ stability.\cite{belik11}

\section*{Acknowledgments}
The authors thank Karin Rabe, Turan Birol, Darrell Schlom and James Rondinelli for helpful discussions. N.\ A.\ B.\ and C.\ J.\ F.\ were supported by DOE-BES under Award Number DE-SCOO02334. 

\providecommand*\mcitethebibliography{\thebibliography}
\csname @ifundefined\endcsname{endmcitethebibliography}
  {\let\endmcitethebibliography\endthebibliography}{}

\end{document}